\documentclass[aps,nofootinbib,twocolumn,floatfix,superscriptaddress,secnumarabic]{revtex4-1}
\usepackage{amsfonts,amssymb,epsfig,verbatim,mathbbol,amstext,amsmath,psfrag}
\usepackage{graphicx}
\usepackage{t1enc}
\usepackage{amsmath}
\usepackage{subfigure}
\usepackage{dsfont}
\usepackage{slashed}

\usepackage{colordvi}
\usepackage{xcolor}
\usepackage{color}

\newif\ifSMver
\SMvertrue

\ifSMver

\newcommand{\appendixsection}{%
\setcounter{equation}{0}
\setcounter{section}{0}
\renewcommand{\theequation}{S\arabic{equation}}
\onecolumngrid
\vspace*{.7cm}
\hrule
\vspace*{.04cm}
\hrule
\begin{center}
\vspace*{.4cm}
{\bf \large Supplemental Material}
\vspace*{.5cm}
\end{center}
\twocolumngrid
}
\else

\newcommand{\appendixsection}{\appendix}
\fi

\newcommand{\dd}{\textmd{d}}
\newcommand{\be}{\begin{equation}}
\newcommand{\ee}{\end{equation}}

\newcommand{\expv}[1]{\left \langle #1 \right \rangle}
\newcommand{\tr}{\textmd{tr}\,}

\newcommand{\mpi}{M_{\pi}}
\newcommand{\fpi}{f_{\pi}}
\newcommand{\fpip}{f'_{\pi}}
\newcommand{\fpipp}{f''_{\pi}}

\usepackage[    bookmarks,
                 bookmarksopen = true,
                 bookmarksnumbered = true,
                 breaklinks = true,
                 linktocpage,
                 colorlinks = true,
                 linkcolor = blue,
                 urlcolor  = blue,
                 citecolor = blue,
                 anchorcolor = green,
                 hyperindex = true,
                 hyperfigures]
		{hyperref}        
		
\begin{document}
\title{Weak decay of magnetized pions}

\author{G.~S.~Bali}
\affiliation{Institute for Theoretical Physics, Universit\"at Regensburg, D-93040 Regensburg, Germany}
\affiliation{Department of Theoretical Physics, Tata Institute of Fundamental Research, Homi Bhabha Road, Mumbai 400005, India.}
\author{B.~B.~Brandt}
\affiliation{Institute for Theoretical Physics, Goethe Universit\"at Frankfurt, D-60438 Frankfurt am Main, Germany}
\author{G.~Endr\H{o}di}
\affiliation{Institute for Theoretical Physics, Goethe Universit\"at Frankfurt, D-60438 Frankfurt am Main, Germany}
\author{B.~Gl{\"a}{\ss}le}
\affiliation{Zentrum f\"ur Datenverarbeitung (ZDV), Universit\"at T\"ubingen, W\"achterstr.\ 76, D-72074 T\"ubingen, Germany}

\begin{abstract}
The leptonic decay of charged pions 
is investigated in the presence of background magnetic fields.
In this situation Lorentz symmetry is broken and new fundamental
  decay constants need to be introduced, associated with the decay via the 
  vector part of the electroweak current.
We calculate the magnetic field-dependence of both the usual and a
new decay constant non-perturbatively on the lattice. We employ both Wilson and staggered quarks and extrapolate the results to the continuum limit. 
With this non-perturbative input we calculate the tree-level 
electroweak amplitude for the full decay rate in strong magnetic fields. 
We find that the muonic decay of the charged pion is enhanced drastically by the magnetic field.
We comment on possible astrophysical implications.
\end{abstract}


\maketitle 

{\em Introduction}.
Strong (electro)magnetic fields bear a significant impact on the physics of various systems 
ranging from off-central 
heavy-ion collisions through the evolution of the early universe to magnetized neutron stars (magnetars). 
In particular,
many novel phenomena emerge from the competition between electromagnetism
and color interactions if the magnetic field $B$ becomes similar in
magnitude to the strong interaction scale: $eB\sim\Lambda^2_{\rm QCD}$.
If the time-scale of the fluctuations 
in $B$ is larger than other relevant scales of the problem, it is 
reasonable to treat the magnetic field classically as a background field. 
For reviews on this subject, see for example Refs.~\cite{Kharzeev:2013jha,Miransky:2015ava}.

Such a background magnetic field is known, for instance, to
affect the phase diagram of quantum chromodynamics
(QCD)~\cite{Bali:2011qj,Endrodi:2015oba,Andersen:2014xxa}. 
For cold astrophysical environments,
the low-temperature (hadronic) phase of QCD is particularly relevant. 
In this regime, a prime role is played by the lightest hadrons, i.e.\ 
pions and kaons. 
Specifically, their masses appear in the nuclear equation of state within
compact stellar objects and, thus, influence their mass-radius relations.
For stability and equilibrium analyses, the respective decay rates are
equally important. 
Dominant 
cooling mechanisms for magnetars~\cite{Duncan:1992hi} involve (inverse) 
$\beta$-decay,
photo-meson interactions and pion decay~\cite{Waxman:1997ti}. 
Pions radiate 
energy via inverse Compton scattering until they decay, 
imprinting the spectrum of the 
subsequently produced neutrinos~\cite{Zhang:2002xv}.
Strong electromagnetic fields are also created in violent astrophysical
processes 
such as neutron star mergers and supernova events, where 
weak nuclear reactions and decays govern cooling mechanisms and affect 
the neutrino spectrum~\cite{Metzger:2018uni}.

The $B$-dependence of pion masses has been investigated 
in various settings, ranging from chiral perturbation theory~\cite{Agasian:2001ym,Andersen:2012zc}
through numerical lattice QCD simulations~\cite{Bali:2011qj,Hidaka:2012mz,Luschevskaya:2014lga,Bali:2017ian} to 
model approaches~\cite{Fayazbakhsh:2012vr,Avancini:2016fgq,Zhang:2016qrl,Mao:2017wmq,GomezDumm:2017jij,Coppola:2018vkw}.
Less is known about the decay rates for nonzero magnetic fields. 
The decay constant for the neutral pion has been studied
in chiral perturbation theory~\cite{Agasian:2001ym,Andersen:2012zc,Andersen:2012dz}
and in model settings~\cite{Fayazbakhsh:2012vr,Fayazbakhsh:2013cha,Avancini:2016fgq,Zhang:2016qrl,Mao:2017wmq,GomezDumm:2017jij}. The decay constant of the charged pion has only been discussed 
so far in chiral perturbation 
theory~\cite{Andersen:2012zc}. 

In this letter we investigate the magnetic field-dependence of the decay 
rate of charged pions at zero temperature. We demonstrate that the previous studies in this
direction are incomplete: in the presence of the magnetic field both neutral and charged pions
have two 
independent decay constants, 
of which only one has been investigated up to now. 
We determine both decay constants for charged pions non-perturbatively on the 
lattice, employing two different fermionic discretizations.  
Using this QCD input, we proceed to calculate the weak decay rate 
using leading-order electroweak
perturbation theory. For this calculation we employ the 
lowest Landau level (LLL) approximation for the outgoing charged lepton state,
which is a viable simplification for strong background magnetic fields. 
Our preliminary results using Wilson fermions on a reduced set of lattice spacings
were presented in Ref.~\cite{Bali:2017yku}.

{\em Pion decay constants}.
The pion decay constant is related to the hadronic matrix elements $H_\mu$ of the weak interaction 
current between the vacuum and a pion state with momentum $p_\mu$.
For $B=0$, parity dictates that the matrix element
$\langle 0|\bar{d}\gamma_{\mu} u|\pi^-\rangle$
vanishes, since the only Lorentz-structure available
is $p_\mu$:
\be
\begin{split}
H_\mu&\equiv\big\langle0|\bar d(x) \gamma_\mu(1-\gamma^5) u (x)| \pi^-(p)\big\rangle \\
&=\!-\expv{0|\bar d(x) \gamma_\mu\gamma^5 u (x)| \pi^-(p)} 
= \!-i e^{ipx}\, \fpi p_\mu\,.
\end{split}
\label{eq:Fdef}
\ee
The coefficient $\fpi$ is the pion decay constant, which coincides for 
negatively and positively charged pions due to charge conjugation symmetry.
Throughout this letter we use the normalization where $\fpi\approx131\textmd{ MeV}$
for a physical pion in the vacuum.

In the presence of a background electromagnetic field $F_{\mu\nu}$ the relation~\eqref{eq:Fdef} 
takes a more general form. Exploiting Lorentz-covariance, using the tensor $F_{\mu\nu}$ and the vector
$p_\mu$, additional vector and axial vector combinations can be formed:
\be
\begin{split}
\!\!\!\big\langle0|\bar d(x) \gamma_\mu \gamma^5 u (x)| \pi^-(p)\big\rangle &= i e^{ipx} 
\bigg[ \fpi p_\mu + \fpipp eF_{\mu\nu}p^\nu
\bigg]\,,\\
\big\langle0|\bar d(x) \gamma_\mu  u (x)| \pi^-(p)\big\rangle &= i e^{ipx} 
\bigg[ i\frac{\fpip}{2} \epsilon_{\mu\nu\rho\sigma} eF^{\nu\rho}p^\sigma 
\bigg]\,,
\end{split}
\label{eq:Fdef2}
\ee
where $e$ denotes the elementary charge and we follow the convention 
$\epsilon^{0123}=+1$. 
Charge conjugation implies that the decay rate 
is the same for positively and negatively 
charged pions 
and is also independent of the direction of the magnetic field.
This is ensured by 
the ratios $\fpi/\fpip$ and $\fpi/\fpipp$ being real, as we will see below.
In our conventions, all three decay constants are real and positive. 
We remark that the new Lorentz structures also
exist for matrix elements involving neutral pions.

We consider a background magnetic field 
$\mathbf{B}$ that points in the $z$ direction, 
implying $F_{21}=-F_{12}=B$. 
For a pion of mass $\mpi$ with vanishing momentum along the magnetic field, $p_3=0$,
\be
H_0  
= 
-ie^{i\mpi x_0}  \fpi \mpi, \quad
H_3 = e^{i\mpi x_0} \fpip \,eB \mpi\,. 
\label{eq:Dmu}
\ee
For charged states that are in
the LLL, only these two components of $H_\mu$ contribute to the decay rate. For
details on this and further elements of the perturbative calculation, we refer to 
Secs.~\ref{app:LLL} and~\ref{sec:appdr} of the Supplemental Material.
The decay constants $\fpi$ and $\fpip$ 
depend on the Lorentz-scalars
$F_{\mu\nu}F^{\mu\nu}/2=B^2$ and $p_\mu p^\mu=\mpi^2(B)$. 

The matrix element of the vector current can also be
interpreted 
from a different perspective: the magnetic field mixes the pion 
with the $\rho$-meson having zero
spin projection along the magnetic field (i.e.\ $s_3=0$)~\cite{Bali:2017ian}. 
Since the latter has the same quantum numbers as the $\mu=3$ component of the vector part of the electroweak current, this mixing 
gives rise to a nonzero 
value for the vector matrix element, the second relation of Eq.~\eqref{eq:Fdef2}.

We mention that for nonzero temperature an additional vector $u_\mu$ describing the thermal medium
($u_0\neq u_i$) appears and leads to 
a splitting between spatial and temporal decay constants (see, for example, 
Ref.~\cite{Fayazbakhsh:2013cha}).
Here we work at $T=0$, where this effect is absent. 
Furthermore, note that the presence of the two terms $\fpi$ and $\fpipp$ in the first relation of Eq.~\eqref{eq:Fdef2} implies that 
the axial vector matrix element is different for indices $\mu=0,3$ and 
$\mu=1,2$, as was also found in Ref.~\cite{Fayazbakhsh:2013cha}. 
However, for a purely magnetic background the term involving $\fpipp$ 
is absent from $H_0$ and $H_3$.

{\em Pion decay rate}.
The weak interaction matrix element~\eqref{eq:Dmu} enters the rate 
of the leptonic decay process
$\pi^-(p) \to \ell^-(k) \,\bar\nu_\ell(q)$,
where $p$, $k$ and $q$ denote the four-momenta of the pion, the charged lepton $\ell$ 
and the antineutrino $\bar\nu_\ell$, respectively. 
The decay into a muon $\ell=\mu$ is the dominant channel, 
with a decay fraction of $99.98\%$~\cite{Agashe:2014kda} at $B=0$.

We work at the tree level of electroweak perturbation theory and employ the effective, four-fermion 
interaction with Fermi's constant $G$ as coupling. 
Due to the current-current 
structure of the effective electroweak Lagrangian~\cite{okun2013leptons,schwartz2014quantum},
the decay amplitude factorizes into leptonic and hadronic parts,
$\mathcal{M} = G/\sqrt{2}\cdot \cos\theta_c \,L^\mu
H_\mu$, 
where the Cabibbo angle $\theta_c$ entered due to the 
mixing between the down and strange quark mass eigenstates. 
The relevant hadronic components $H_\mu$ are shown in Eq.~\eqref{eq:Dmu}. 
Moreover, the leptonic component reads
$L^\mu \equiv 
\bar u_\ell(k) \gamma^\mu(1-\gamma^5) v_\nu(q)$
in terms of the bispinor solutions $u_\ell$ and $v_\nu$.

The decay rate $\Gamma$ involves the modulus square of the amplitude,
integrated over the phase space and 
summed over
the intrinsic quantum numbers of the outgoing asymptotic states.
To find the latter for the charged lepton, 
we need the bispinor solutions of the Dirac equation for $B>0$.
These are the so-called Landau levels --- orbits localized 
in the spatial plane perpendicular to $\mathbf{B}$ with quantized radii. 
The Landau levels come with a  
multiplicity proportional to the flux $\Phi=|eB|\cdot L^2$ of the magnetic field. 
In order to regulate this multiplicity, we need to assume
that the outgoing states are defined in a finite spatial volume $V=L^3$. 
For the decay rate such volume factors will cancel. 

For strong fields
the dominant contribution 
stems from the lowest Landau level.
The sum over the multiplicity of the 
LLL states gives~\cite{Bhattacharya:2007vz}
\be
\sum_{\textmd{LLL}} u_\ell(k) \bar u_\ell(k) = 
\frac{\Phi}{2\pi}\cdot
(\slashed{k}_\parallel +m_\ell) \cdot \frac{1-\sigma^{12}}{2}  \,,
\label{eq:spinsumLLL}
\ee
where $\slashed{k}_\parallel=k^0\gamma^0-k^3\gamma^3$ and $\sigma^{12}=i\gamma^1\gamma^2$ is the relativistic spin operator. 
Eq.~\eqref{eq:spinsumLLL} reflects the fact that the LLL solutions have their spin 
anti-aligned with the magnetic field (since the lepton has negative charge) 
and are characterized only by the momentum along the $z$ direction (i.e.\ along $\mathbf{B}$). 
Due to angular momentum conservation the antineutrino spin is also 
aligned with the magnetic field.
Moreover, the right-handedness of the antineutrino also sets 
the direction of its momentum to be parallel to the magnetic field.

Having determined $|\mathcal{M}|^2$, 
we finally need to integrate over the phase space for the outgoing particles. 
The resulting decay rate reads
\be
\Gamma(B)
= 
|eB|\,\frac{G^2}{4\pi}\cos^2\!\theta_c \,|\fpi+i\fpip eB|^2\frac{m_\ell^2}{\mpi}\,.
\label{eq:GB19}
\ee
As anticipated above, the decay rate only depends on the magnitude of $B$,
due to the absence of an interference term in
$|\fpi+i\fpip eB|^2=\fpi^2+[\fpip eB]^2$.
Dividing by the $B=0$ result~\cite{okun2013leptons}, 
the dependence on $G$ and $\theta_c$ cancels:
\be
\frac{\Gamma(B)}{\Gamma(0)} = \frac{\fpi^2+[\fpip eB]^2}{\fpi^2(0)} 
\cdot \left[1-\frac{m_\ell^2}{\mpi^2(0)}\right]^{-2} \!\!\!\cdot\frac{2|eB|}{\mpi(0)\mpi(B)}\,.
\label{eq:Gammarat}
\ee
We stress that this result was obtained using the LLL
approximation, which is in general valid for strong fields~\cite{Miransky:2015ava,Bruckmann:2017pft}. 
For the leading-order perturbative decay rate, 
higher Landau-levels turn out to give zero contribution for 
$eB> \mpi^2(0) -m_\ell^2$.

{\em Lattice setup}.
Eq.~\eqref{eq:Gammarat} contains three non-perturbative parameters, which describe the response
of the pion to the background field: $\mpi(B)$, $\fpi(B)$ and $\fpip(B)$. 
We calculate these via two independent sets of lattice QCD simulations. First, we work 
with
quenched Wilson quarks. The zero-temperature 
ensembles generated and analyzed in Ref.~\cite{Bali:2017ian} are supplemented
by a fourth, finer lattice 
ensemble, so that the lattice spacing spans $0.047\textmd{ fm}\le a\le 0.124\textmd{ fm}$. 
The $B=0$ pion mass is set to $\mpi(0)\approx 415\textmd{ MeV}$. 
To remove $B$-dependent $\mathcal{O}(a^2)$ effects on quark masses, the bare mass
parameters are tuned to fall on the magnetic 
field-dependent line of constant physics determined in Ref.~\cite{Bali:2017ian}. 

In the second set of simulations we work with $N_f=2+1$ flavors of 
dynamical staggered fermions, using the 
ensembles of Refs.~\cite{Bali:2011qj,Bali:2012zg}. The employed lattice spacings
lie in the range $0.1\textmd{ fm}\le a\le 0.22\textmd{ fm}$, and the quark masses are 
set to their physical values~\cite{Borsanyi:2010cj} such that $\mpi(0)\approx 135\textmd{ MeV}$. 
For both formulations we perform a continuum extrapolation based on the available four 
lattice spacings.
This enables us to quantify the systematics related to the differences between the two approaches:
heavier-than-physical versus physical 
pion mass and quenched versus dynamical quarks.
We remark that simulations with
dynamical Wilson quarks in the presence of a background magnetic field
would require computational resources that are by orders of
magnitude larger than those used for the current study.

The general measurement strategy involves the analysis of the 
matrix elements $H_0$ and $H_3$ of Eq.~\eqref{eq:Dmu}. 
These are encoded in the spatially averaged Euclidean correlators
$C_{\mathcal{O}P}(t) = \big\langle\sum_\mathbf{x}\mathcal{O}(\mathbf{x},t) \,P^\dagger(\mathbf{0},0)\big\rangle$
with $\mathcal{O}$ being either of
$P =\bar u \gamma^5 d$, 
$A =\bar u \gamma_0 \gamma^5 d$
or $V =\bar u \gamma_3 d$.
In the large-$t$ limit, the dominant contribution to the spectral representation of all three correlators 
comes from a pion state.
We fit the three correlators using 
\be
C_{\mathcal{O}P}(t) = c_{\mathcal{O}P}
\left[ e^{-\mpi t} \pm e^{-\mpi(N_t-t)} \right]\,,
\label{eq:fit1}
\ee
where the positive sign is 
taken for $C_{PP}$ and the negative for $C_{AP}$ and $C_{V\!P}$ due to the time reversal properties of the 
correlators. The decay constants are extracted via
\be
\fpi = Z_A\cdot \frac{\sqrt{2} \,c_{AP}}{\sqrt{\mpi c_{PP}}}, \quad
i\fpip eB= Z_V\cdot \frac{\sqrt{2}\, c_{VP}}{\sqrt{\mpi c_{PP}}}\,,
\label{eq:fit2}
\ee
where $Z_A$ and $Z_V$ are the multiplicative renormalization constants of the axial vector and 
vector currents. 

For Wilson quarks we employ smeared pseudoscalar sources (for more details, see Ref.~\cite{Bali:2017ian})
and fit all three correlators 
simultaneously. For the staggered 
analysis we work with point sources and fit the $C_{AP}$ and $C_{PP}$ correlators to find $\mpi$ and $\fpi$. 
In a second step, volume sources are employed for $C_{V\!P}/C_{AP}$ to enhance the signal in $\fpip/\fpi$. 
The staggered discretization of $A$ and $V$ requires operators nonlocal in 
Euclidean time and has been worked out in Ref.~\cite{Kilcup:1986dg}. 
For staggered quarks and the currents we use
the renormalization constants are trivial, $Z_A=Z_V=1$. 
For Wilson quarks this is not the case, nevertheless, these ultraviolet quantities
are expected to be independent of the magnetic field. We employ the $B=0$ non-perturbative results of Ref.~\cite{Gimenez:1998ue} 
(see also Ref.~\cite{Gockeler:1998ye}) and fit these in combination with the asymptotic perturbative two-loop results of Ref.~\cite{Skouroupathis:2008mf}
(see also Ref.~\cite{Bali:2013kia}) to a Pad\'e parametrization.

{\em Results}.
Inspecting the $B>0$ correlation functions,
we see clear signals for $C_{V\!P}$ (see Sec.~\ref{sec:corrlat} of the Supplemental Material), which vanishes at $B=0$. 
The mass and the decay constants are extracted using the fits described in Eqs.~\eqref{eq:fit1} and~\eqref{eq:fit2}. For the complete magnetic field range, the pion mass is found to be described within $5\%$
by the formula
\be
\mpi/\mpi(0)=\sqrt{1+|eB|/\mpi^2(0)}\,,
\label{eq:mpiB}
\ee
which assumes pions to be point-like free scalars. 
This has been observed many times in the literature,
both using dynamical staggered~\cite{Bali:2011qj}, quenched Wilson~\cite{Hidaka:2012mz,Bali:2017ian} and quenched overlap quarks~\cite{Luschevskaya:2014lga}.

\begin{figure}[b]
 \centering
 \includegraphics[width=8cm]{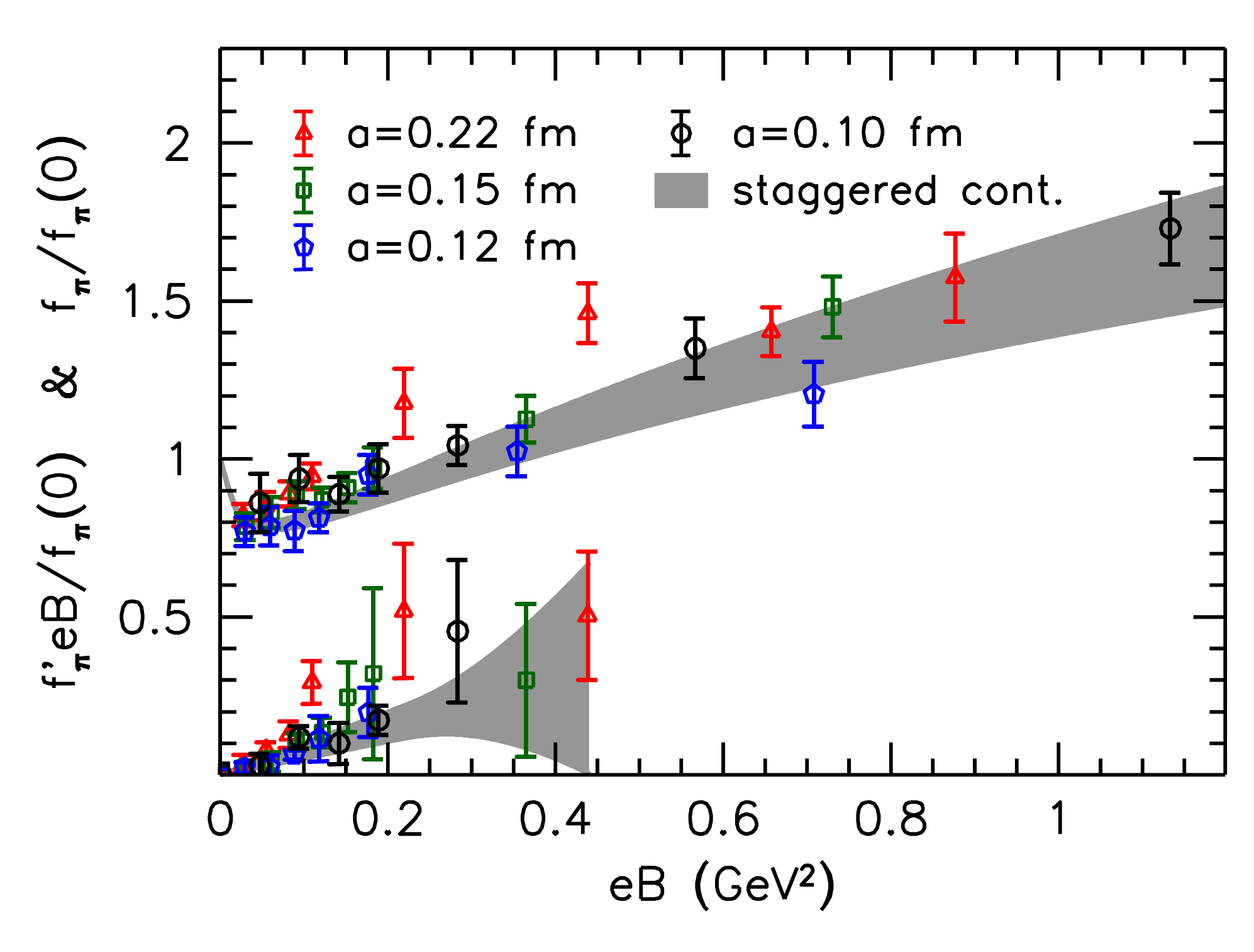}
 \includegraphics[width=8cm]{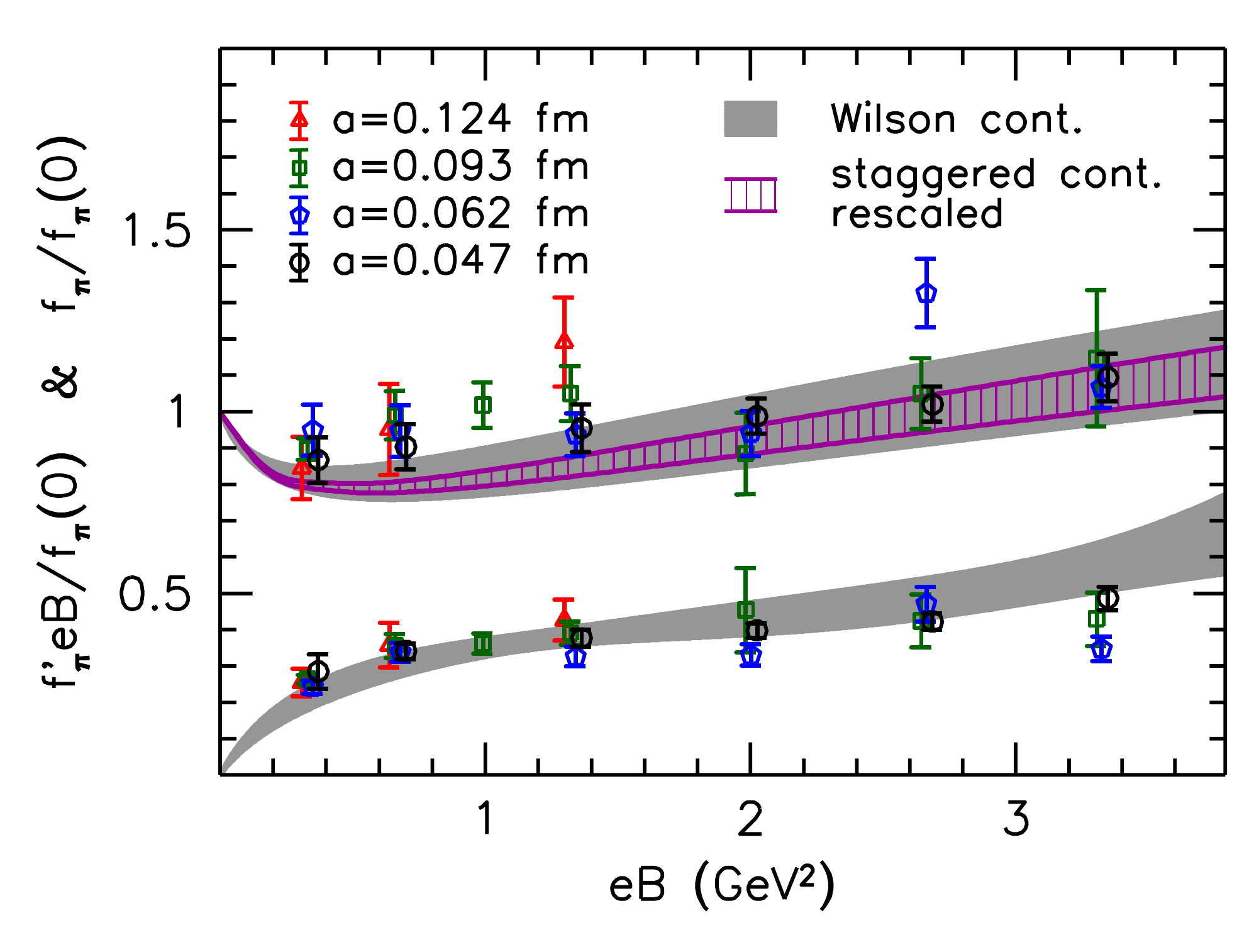}
 \caption{\label{fig:1}Continuum extrapolation (gray bands) of the decay constants for staggered (upper panel)
 and Wilson quarks (lower panel). Both panels include results for $\fpi/\fpi(0)$ (upper points) 
 and for $\fpip eB/\fpi(0)$ (lower points). The staggered results were obtained 
 at the physical point, while the Wilson results correspond to a $B=0$ pion mass of $\mpi(0)=415\textmd{ MeV}$. For $\fpi/\fpi(0)$ we also compare the 
 two continuum extrapolations after a rescaling of the magnetic field for the staggered
 curve (purple band; see the text for details). }
\end{figure}

The normalized combinations $\fpi/\fpi(0)$ and $\fpip eB/\fpi(0)$
are shown for four lattice spacings in Fig.~\ref{fig:1}, both for staggered 
and for Wilson fermions. 
To parameterize the $B$-dependence, we found 
that it is advantageous to consider polynomial fits for the 
amplitudes $\fpi \mpi$ and $\fpip eB \mpi$
of the matrix elements of Eq.~\eqref{eq:Dmu}. The continuum extrapolation is carried out by
including lattice artefacts of $\mathcal{O}(a)$ (for Wilson) and 
$\mathcal{O}(a^2)$ (for staggered) in the coefficients. Specifically, 
the parameterizations of the individual 
decay constants read
\be
\begin{split}
\fpi/\fpi(0) &=  [ 1 + c_1 \,|eB| \,] \cdot \mpi(0)/ \mpi\,, \\
\fpip/\fpi(0) &= [ d_0 + d_1 \,|eB| + d_2 \,|eB|^2 ] \cdot \mpi(0)/ \mpi \,, \\
\end{split}
\ee
and $\mpi/\mpi(0)$ is taken from Eq.~\eqref{eq:mpiB}. 
The quality of the staggered data for $\fpip/\fpi(0)$ only allows
for a fit with $d_1=d_2=0$. For larger magnetic fields we also include a systematic error
estimated using the uncertainties of the data at high $B$.
  Ideally, 
  the analysis in this region should be complemented by additional finer ensembles
  to make the continuum extrapolation of $\fpip$ more robust. 
  Within our range of $B$ fields, however, the decay rate and its
  uncertainty are dominated by $\fpi$.

Motivated by the dependence of $\mpi/\mpi(0)$ on the scaling variable $eB/\mpi^2(0)$, we
compare the continuum extrapolated Wilson results (obtained for $\mpi(0)=415\textmd{ MeV}$) 
to the staggered data (obtained for physical pion masses), after
rescaling the magnetic field for the latter. In particular, we take the staggered 
results for $\fpi/\fpi(0)$ at the magnetic field $eB\cdot (415/135)^2$. 
The resulting curve is also included in the lower panel of Fig.~\ref{fig:1}, 
revealing nice agreement between the two approaches. 
In particular, the slope at the origin is found to be
$-16.9(3) \textmd{ GeV}^{-2}$ 
for staggered and 
$-1.7(2) \textmd{ GeV}^{-2}$ 
for Wilson --- the ratio of which is consistent with the squared pion mass ratio.

For low magnetic fields the ratio 
$\fpip/\fpi(0)$ approaches a constant so that in this case 
the two discretizations can be compared to each other without a similar rescaling. 
We indeed find consistent results: $\fpip/\fpi(0) = 0.8(2) \textmd{ GeV}^{-2}$ for staggered
and $1.2(3) \textmd{ GeV}^{-2}$ for Wilson, respectively. We note that the errors
of the staggered data for this decay constant increase quickly as $B$ grows, rendering 
a comparison for higher magnetic fields inconclusive.
We mention moreover that due to the different treatment of sea quark loops in the 
two approaches, the observed agreement is rather surprising and calls for a better understanding 
of the role of dynamical quarks in the Wilson setup.

\begin{figure}[t]
 \centering
 \includegraphics[width=8cm]{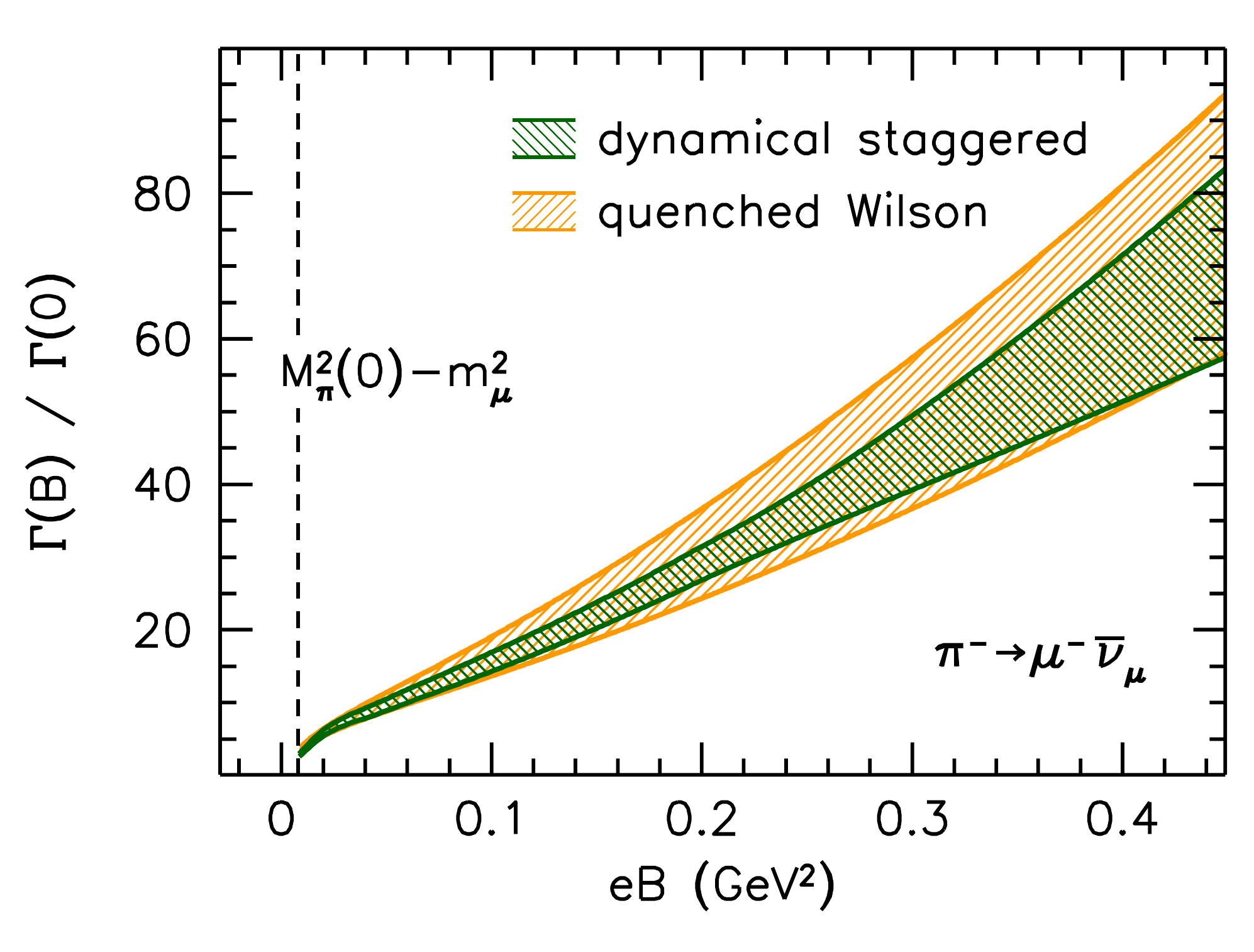}
 \caption{\label{fig:3}The muonic decay rate in units of its $B=0$ value 
 using the continuum extrapolated staggered results with physical quark masses 
 (green). For comparison, 
 the continuum extrapolated Wilson data at higher-than-physical quark masses 
 are also included after a rescaling of the magnetic field by the squared pion 
 mass (yellow). The LLL approximation we employed for the decay rate 
 is valid for $eB>\mpi^2(0)-m_\mu^2$, marked by the dashed vertical line.}
\end{figure}

To determine the decay rate~\eqref{eq:Gammarat} 
we employ the continuum extrapolated staggered results. 
On the basis of the above comparisons, we also 
consider the Wilson results, using a rescaling to the physical point as explained above. 
For the pion mass we use the analytic dependence~\eqref{eq:mpiB}, including a $5\%$ 
systematic error. 
The so obtained curves for the muonic decay rate are shown in Fig.~\ref{fig:3}
for magnetic fields $eB\le0.45\textmd{ GeV}^2$, where 
both staggered and (rescaled) Wilson results are available. 
The decay rate is enhanced drastically by the magnetic field: 
for $eB\approx0.3\textmd{ GeV}^2$ we observe an almost fifty-fold
increase with respect to $B=0$. 
We note that while the ordinary decay mechanism dominates in our study,
the contribution of the new vector decay constant $\fpip$ grows to about $10\%$ 
of the total decay rate at the largest magnetic field of Fig.~\ref{fig:3}.

We remark that
Eq.~\eqref{eq:Gammarat}, supplemented by our staggered lattice results,
suggests the decay rate $K^-\rightarrow \mu^-\bar{\nu}_{\mu}$
to be enhanced only by a factor of
about two at $eB\approx 0.3\textmd{ GeV}^2$. This is mainly due to the
larger mass of the kaon.
Finally, the pion decay rate into electrons
undergoes an enhancement by a factor of about ten. In fact, according to 
Eq.~\eqref{eq:GB19} the ratio of muonic and electronic decay rates
becomes independent of the magnetic field,
\be
\Gamma(B)_{\pi\to e \bar\nu_e}\big/
\Gamma(B)_{\pi\to \mu \bar\nu_\mu}
= 
(m_e/m_\mu)^2
\approx 2.27\cdot 10^{-5},
\label{eq:Gemurat}
\ee
and is by about a factor of $5.4$ smaller than the corresponding
fraction $1.23\cdot 10^{-4}$ at $B=0$.

{\em Conclusions}.
In this letter we computed the rate for the leptonic decay of charged pions in the presence 
of strong background magnetic fields. The result is given by Eq.~\eqref{eq:Gammarat}, for which 
we employed electroweak perturbation theory and the lowest-Landau-level approximation for 
the outgoing charged lepton $\ell$, valid for strong fields. Including higher-order
terms in the electroweak calculation (see, e.g., Refs.~\cite{Lubicz:2016xro,Patella:2017fgk}), as well as going beyond the lowest Landau level
is possible, allowing to systematically improve this result.
In this case also the constant $\fpipp$, that we have not
determined here, may enter.

We demonstrated that --- besides the ordinary pion decay constant $\fpi$ --- 
the decay rate depends on an additional fundamental parameter $\fpip$. The latter 
decay constant characterizes a new decay mechanism
that becomes available 
for nonzero magnetic fields. We calculated both decay constants, together with the pion mass,
using lattice simulations employing dynamical staggered quarks with physical masses,
and also compared to the results of quenched Wilson simulations with heavier-than-physical
quarks. For both cases, continuum extrapolations were carried out to eliminate lattice 
discretization errors. For low magnetic fields, we obtain for the new decay constant 
$\fpip=0.10(2)\textmd{ GeV}^{-1}+\mathcal{O}(B)$.

Our final result for the full decay rate is visualized in Fig.~\ref{fig:3}, revealing 
a dramatic enhancement of the rate or, correspondingly, a drastic 
reduction of the mean lifetime $\tau_\pi=1/\Gamma$. 
A typical $B>0$ lifetime is
\begin{equation*}
\tau_\pi=5\cdot10^{-10}\textmd{ s}\quad\,\textmd{for}\quad\,
B\approx 0.3\textmd{ GeV}^2/e \approx 5\cdot10^{15}\textmd{ T} \,.
\end{equation*}
Since lifetimes of magnetic fields in off-central heavy-ion collisions
are by $14-15$ orders of magnitude smaller~\cite{Kharzeev:2013jha}, 
it is clear that this effect will not result in any observable
predictions for heavy-ion phenomenology. 
However, the $B$-dependence of weak decays is expected to be essential in astrophysical environments. (Notice that the upper limit for magnetic field strengths in the core
  of magnetized neutron stars is thought to be around $B=10^{14}-10^{16}\textmd{ T}$~\cite{1991ApJ...383..745L,Ferrer:2010wz}.)
Indeed, for $B=0$ the pion mean lifetime and the time-scale  for cooling
via inverse Compton scattering are roughly comparable~\cite{Zhang:2002xv}. 
Thus, a reduction in $\tau_\pi$ will inevitably decrease
radiation energy loss of pions and result in a harder neutrino spectrum.

Similarly to the pion decay rate, the magnetic field will have 
an impact on (inverse) $\beta$-decay rates 
and nucleon electroweak transition form-factors.
Indeed, an enhancement by the magnetic field is expected 
for processes involving nucleons as well~\cite{Matese:1969zz,FassioCanuto:1970wk,Shinkevich:2004ja}, 
see, e.g., the review~\cite{Giunti:2014ixa}.
The tools developed in the present letter will also be useful 
to study these effects that are relevant
for cold and magnetized environments.\\

\noindent
{\em Acknowledgments}.
This research was funded by the DFG (Emmy Noether Programme EN 1064/2-1 and
SFB/TRR 55). GB thanks Basudeb Dasgupta for discussion. The computations
were carried out on the GPU, iDataCool and Athene~2 clusters
of Universit\"at Regensburg.

\appendixsection

\section{Dirac bispinors for \boldmath $B>0$}
\label{app:LLL}

We work with the metric $g^{\mu\nu}=\textmd{diag}(1,-1,-1,-1)$ and use Minkowskian Dirac matrices with the 
conventions $\sigma^{12}=i\gamma^1\gamma^2$ and $\gamma^5=i\gamma^0\gamma^1\gamma^2\gamma^3$. 

To write down the bispinor states for the outgoing charged lepton,
we need to solve the Dirac equation for $B>0$. 
The solutions are the so-called Landau levels, indexed by the Landau index $n$. 
These have the energies
\be
E_{n,k_3,s_3} = \pm \sqrt{(2n+1+2s_3) eB +k_3^2 + m_\ell^2}\,,
\label{eq:LLenergies}
\ee
where $k_3$ is the momentum along the magnetic field --- which is chosen to lie 
in the positive $z$ direction --- and $s_3$ is the spin projection on the $z$ axis. 
Note that the lowest energy is $n=k_3=0$ with $s_3=-1/2$, reflecting the fact 
that the lepton spin (due to its negative charge $-e$) tends to align itself
anti-parallel to $\mathbf{B}$. Higher-lying levels have energies of at least $\sqrt{2eB}$
and are in general not expected to be relevant at low temperatures and 
densities,
as long as $eB\gg m_{\ell}^2$.
Here we employ the lowest Landau level (LLL) approximation ($n=0$, $s_3=-1/2$), which simplifies the 
discussion considerably. The calculation can be generalized
to take into account all levels. 

We start from the LLL modes calculated in 
Ref.~\cite{Bhattacharya:2007vz} in the gauge $A_1=-Bx_2$ and in the infinite volume
(see also Ref.~\cite{Miransky:2015ava}). 
Here we consider the volume to be large but finite to regulate the number of modes.  
In a volume $V=L^3$ this degeneracy factor is 
$\Phi/(2\pi)$ with $\Phi=|eB| L^2$ being
the flux of the magnetic field through the area of the system.
The modes in coordinate space read,\footnote{Compared to Ref.~\cite{Bhattacharya:2007vz} we insert a factor $\sqrt{k_0+m_\ell}\cdot\sqrt{L}$ for a more
  convenient normalization.}
\be
\begin{split}
u_\ell(k) = \!\!&\,\,\sqrt{k_0+m_\ell}\cdot\left(\frac{eBL^2}{\pi}\right)^{\!1/4}\cdot
 \left(0, 1, 0, \frac{-k_3}{k_0+m_\ell}\right)^{\!\intercal}
 \\
 &\cdot\exp\!\left[ik_0x_0-ik_1x_1 -ik_3x_3-\frac{eB}{2}\!\left(x_2+\frac{k_1}{eB}\right)^{\!\!2} \right],
\end{split}
 \label{eq:eminusk}
\ee
where the energy is $k_0\equiv E_{0,k_3,-1/2}=\sqrt{k_3^2+m_\ell^2}$. Note that while $k_3$ is a momentum, $k_1$ is a 
quantum number indexing the degeneracy of the Landau-modes and has no momentum interpretation. 
Also note that the LLL can only have negative spin so that
\be
\sigma^{12} u_\ell(k) = -u_\ell(k)\,.
\ee
The normalization of the state~\eqref{eq:eminusk} is
\be
\int\! \dd^3\! x \,u_\ell^\dagger(p) \,u_\ell(k) = (2\pi)^2 \delta^{(2)}_{1,3}(p-k) \,2k_0\,L\,.
\ee
Thus, the states with different $k_1$ quantum numbers are orthogonal to each other. Regulating the $\delta$-functions in a finite volume\footnote{The Fourier-transform of the $\delta$-function in a finite linear size $L$ 
is
$(2\pi)\delta(p) = \int\! \dd x\, e^{ipx}$ so that $\delta(0)=L/(2\pi)$.}, we
have
\be
|u_\ell(k)|^2=\int\! \dd^3\! x \,u_\ell^\dagger(k)\,u_\ell(k) = 2k_0V \,,
\label{eq:normalization}
\ee
which coincides with the usual normalization of bispinors (cf.\ Ref.~\cite{schwartz2014quantum}).

Having specified the normalization of the charged lepton state, we can calculate the 
sum over the intrinsic quantum numbers. For our LLL states this does not 
involve a spin-sum, since only $s_3=-1/2$ is allowed. However, we have to 
sum over the unspecified quantum 
number $k_1$.
In a finite volume, this index takes $\Phi/(2\pi)$ different values
and can be approximated as $L/(2\pi)$ times the integral over $k_1$,
\be
\sum_{\textmd{LLL}} \equiv \sum_{k_1} = L\int \!\frac{\dd k_1}{2\pi}\,.
\ee
After some algebra we obtain (see also Ref.~\cite{Bhattacharya:2007vz})
the result that we quoted in 
Eq.~(4) in the body of the text.

\section{Decay rate for \boldmath$B>0$}
\label{sec:appdr}

As explained in the main text, the antineutrino has fixed spin 
(in the positive $z$ direction) as well as fixed chirality. 
The corresponding bispinor solution $v_\nu(q)$ necessarily has 
vanishing perpendicular momenta $q_1=q_2=0$,  satisfies the on-shell relation 
$q_0=|q_3|$ and fulfills
\be
v_\nu(q) \bar v_\nu(q) = P_\sigma \,\slashed{q}\,  P_\sigma, \quad\quad P_\sigma=\frac{1-\sigma^{12}}{2}\,,
\label{eq:neutrinospin}
\ee
keeping in mind that the physical spin for antiparticles is the 
opposite of the eigenvalue of $P_\sigma$. 
Inserting~\eqref{eq:neutrinospin} and the spin sum~(4)
into the modulus square of the summed amplitude, we obtain
\be
\mathcal{A} = \sum_{\textmd{LLL}} \mathcal{M}\mathcal{M}^*
= \frac{G^2 \,\Phi}{4\pi}\cos^2\!\theta_c  \,H_\mu H^*_\rho \,\mathcal{T}^{\mu\rho}\,,
\label{eq:A0}
\ee
where 
\be
\mathcal{T}^{\mu\rho}\equiv
\tr \!\!\left[ P_\sigma \slashed{q}P_\sigma \gamma^\rho (1-\gamma^5) (\slashed{k}_\parallel +m) P_\sigma \gamma^\mu (1-\gamma^5) \right]\,.
\ee
The trace is simplified using standard identities, revealing 
that $\mathcal{T}^{\mu\rho}$ is only nonzero if the indices $\mu$ and $\rho$ 
assume the values $0$ or $3$. Thus, only $H_0$ and $H_3$ contribute to 
the decay rate, as anticipated in the main text. 
Making use of Eq.~(3) we arrive at
\be
\mathcal{A} = \frac{G^2\,\Phi}{\pi}\cos^2\!\theta_c \,\mpi^2 \,(k^0+k^3)(q^0+q^3) 
\, |\fpi+i\fpip eB|^2\,.
\label{eq:A1}
\ee

We proceed by specifying the kinematics in the frame where the 
pion has vanishing momentum along the magnetic field (and is in the LLL). 
Together with the conservation rules, the on-shell conditions for the outgoing leptons
fully specify the magnitude of $q^3=-k^3$ in terms of the masses $\mpi$ and $m_\ell$,
the decay only being possible for $\mpi>m_\ell$. 
Due to the right-handedness of the antineutrino and the fixed spins of the leptons, 
$k^3$ must in fact be negative, also visible 
from the $q^0+q^3=|q^3|+q^3=|k^3|-k^3$ factor of Eq.~(\ref{eq:A1}). 
Thus we have
\be
k^0 = \frac{\mpi^2+m_\ell^2}{2\mpi},\quad\quad k^3=\frac{m_\ell^2-\mpi^2}{2\mpi}\,.
\ee
Inserting this into Eq.~\eqref{eq:A1}, we obtain
\be
\mathcal{A}=\frac{G^2\,\Phi}{\pi}\cos^2\!\theta_c \, |\fpi+i\fpip eB|^2
 (\mpi^2-m_\ell^2) m_\ell^2\,.
 \label{eq:A2}
\ee

Finally, we need the phase space integration factor to obtain the full decay rate $\Gamma$. 
This is the integral of the differential probability $\dd P$ over a time interval $T$~\cite{schwartz2014quantum},
\be
\Gamma=\frac{1}{T}\int\! \dd P\,,
\ee
where, working with $k_3=-k^3>0$,
\be
\begin{split}
\dd P &= \mathcal{A} \cdot \frac{1}{2\mpi V} 
\frac{1}{2q_0 V}\frac{1}{2k_0 V} \cdot \frac{V}{(2\pi)^3}\dd^3q\, \frac{L}{2\pi}\dd k_3 \\
&\hspace*{12pt}\cdot TV (2\pi)^4\delta(\mpi-q_0-k_0) \delta(q_3+k_3) \delta(q_1)\delta(q_2)\,.
\end{split}
\ee
Here the normalization of the one-particle states (for the charged lepton, see Eq.~\eqref{eq:normalization}) was used, and the $\delta$-functions ensure 
that the perpendicular momenta of the antineutrino vanish.\footnote{We note that the same expression can also be obtained by starting with an incoming pion wave function in the LLL (analogous to the spacetime-dependence of Eq.~\eqref{eq:eminusk}), expanding it in plane waves and finding the contribution of each plane wave to $\mathcal{M}$. Care must be taken in this case to normalize the $\delta$-function for the perpendicular antineutrino momenta with the correct $B\to0$ limit.}
Note that the phase space for the charged lepton is  
one-dimensional due to its LLL nature. 
Performing the integral over $q$ and expressing
the energies through $k_3$
we obtain
\be
\Gamma \!=\! 
\frac{\mathcal{A}}{4\mpi L^2} \!\int_{0}^{\infty}\!\! \frac{\dd k_3}{k_3\sqrt{k_3^2+m_\ell^2}} \,\delta\left(\!\mpi-k_3-\sqrt{k_3^2+m_\ell^2}\right) .
\ee
After the variable substitution
$y=k_3+\sqrt{k_3^2+m_\ell^2}$, we insert $\mathcal{A}$ from Eq.~\eqref{eq:A2}, 
to finally arrive at Eq.~(5) of the main text.\footnote{Our 
preliminary result in Ref.~\cite{Bali:2017yku} was missing a factor $1/2$. 
}
As anticipated, 
all volume factors have canceled. 
We note that while the inclusion of higher Landau levels is considerably more 
involved, their contribution is constrained via energy conservation.
Indeed, for the charged lepton to be created with quantum numbers $n$ and $s_3$,
it is required that $\mpi^2>E_{n,0,s_3}^2=(2n+1+2s_3)eB+m_\ell^2$, 
see Eq.~\eqref{eq:LLenergies}. 
Thus, for magnetic fields $eB>\mpi^2(0)-m_\ell^2$, only the LLL ($n=0$, $s_3=-1/2$)
can contribute. For higher orders in electroweak perturbation
theory soft photons become relevant and this simple picture 
changes.

The decay rate may be compared to the $B=0$ result~\cite{okun2013leptons},
\be
\Gamma(0) = \frac{G^2}{8\pi} \cos^2\!\theta_c \,\fpi^2(0)\, [\mpi^2(0)-m_\ell^2]^2\frac{m_\ell^2}{\mpi^3(0)}\,,
\label{eq:GB20}
\ee
where we explicitly indicated that $\mpi$ and $\fpi$ are to be understood at $B=0$.
The ratio of Eqs.~(5) and~\eqref{eq:GB20}
gives Eq.~(6)
in the body of the text. 
For charged kaon decay the calculation is completely analogous, only 
the substitutions $\mpi\to M_K$, $\fpi\to f_K$, $\fpip\to f'_K$ and $\cos\theta_c\to\sin\theta_c$ have 
to be made. 

\section{Correlators on the lattice}
\label{sec:corrlat}

\begin{figure}[b]
 \vspace*{-.3cm}
 \includegraphics[width=8cm]{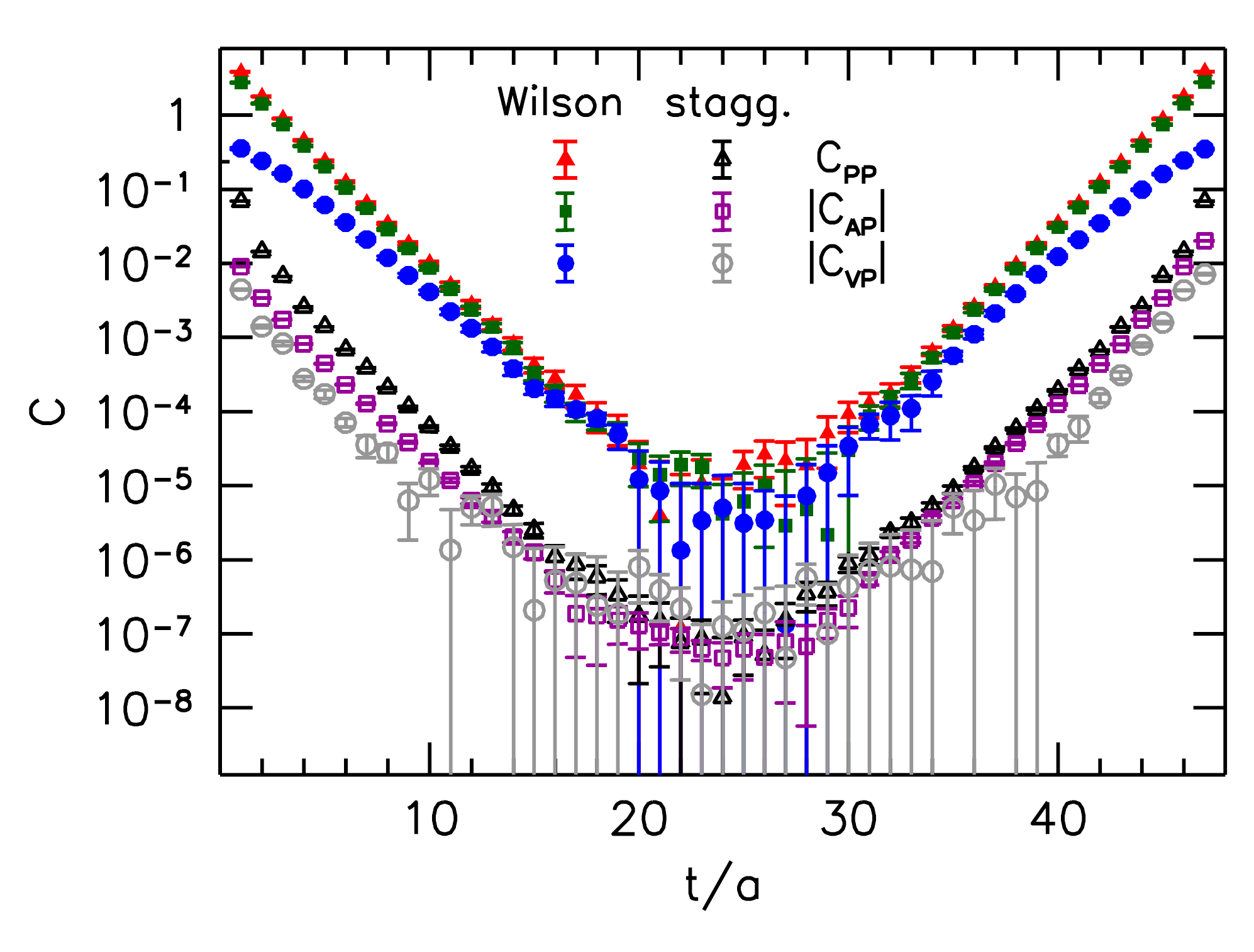}
 \vspace*{-.4cm}
 \caption{\label{fig:correlators} The
 correlators using
 Wilson (upper points) 
 and staggered (lower points) quarks. 
 The employed magnetic fields are $eB\approx 1.33\textmd{ GeV}^2$ (Wilson) 
 and $eB\approx 0.37 \textmd{ GeV}^2$ (staggered).
 For the 
 axial vector and vector operators the absolute value of the correlator is shown.
 The Wilson data is shifted vertically for better visibility.
 }
\end{figure}

The non-perturbative determination of the decay constants involves the analysis 
of the correlators $C_{PP}$, $C_{AP}$ and $C_{V\!P}$ defined in the main text. These are plotted 
in Fig.~\ref{fig:correlators} 
for both fermion formulations for a high magnetic field and meson masses approximately equal 
in lattice units (for the staggered case a kaon correlator is shown). A clear signal is visible for the $C_{V\!P}$ correlator, which would vanish 
on average at $B=0$.

\bibliographystyle{utphys}
\bibliography{fpiB}

\providecommand{\href}[2]{#2}\begingroup\raggedright\begin{thebibliography}{10}

\bibitem{Kharzeev:2013jha}
D.~Kharzeev, K.~Landsteiner, A.~Schmitt, and H.-U. Yee, ``{Strongly Interacting
  Matter in Magnetic Fields},''
\href{http://dx.doi.org/10.1007/978-3-642-37305-3}{{\em Lect. Notes Phys.}
  {\bfseries 871} (2013) pp.1--624}.

\bibitem{Miransky:2015ava}
V.~A. Miransky and I.~A. Shovkovy, ``{Quantum field theory in a magnetic field:
  From quantum chromodynamics to graphene and Dirac semimetals},''
  \href{http://dx.doi.org/10.1016/j.physrep.2015.02.003}{{\em Phys. Rept.}
  {\bfseries 576} (2015) 1--209},
\href{http://arxiv.org/abs/1503.00732}{{\ttfamily arXiv:1503.00732 [hep-ph]}}.

\bibitem{Bali:2011qj}
G.~Bali, F.~Bruckmann, G.~Endr\H{o}di, Z.~Fodor, S.~Katz, {\em et~al.}, ``{The
  QCD phase diagram for external magnetic fields},''
  \href{http://dx.doi.org/10.1007/JHEP02(2012)044}{{\em JHEP} {\bfseries 1202}
  (2012) 044},
\href{http://arxiv.org/abs/1111.4956}{{\ttfamily arXiv:1111.4956 [hep-lat]}}.

\bibitem{Endrodi:2015oba}
G.~Endr\H{o}di, ``{Critical point in the QCD phase diagram for extremely strong
  background magnetic fields},''
  \href{http://dx.doi.org/10.1007/JHEP07(2015)173}{{\em JHEP} {\bfseries 07}
  (2015) 173},
\href{http://arxiv.org/abs/1504.08280}{{\ttfamily arXiv:1504.08280 [hep-lat]}}.

\bibitem{Andersen:2014xxa}
J.~O. Andersen, W.~R. Naylor, and A.~Tranberg, ``{Phase diagram of QCD in a
  magnetic field: A review},''
  \href{http://dx.doi.org/10.1103/RevModPhys.88.025001}{{\em Rev. Mod. Phys.}
  {\bfseries 88} (2016) 025001},
\href{http://arxiv.org/abs/1411.7176}{{\ttfamily arXiv:1411.7176 [hep-ph]}}.

\bibitem{Duncan:1992hi}
R.~C. Duncan and C.~Thompson, ``{Formation of very strongly magnetized neutron
  stars - implications for gamma-ray bursts},''
{\em Astrophys. J.} {\bfseries 392} (1992) L9.

\bibitem{Waxman:1997ti}
E.~Waxman and J.~N. Bahcall, ``{High-energy neutrinos from cosmological
  gamma-ray burst fireballs},''
  \href{http://dx.doi.org/10.1103/PhysRevLett.78.2292}{{\em Phys. Rev. Lett.}
  {\bfseries 78} (1997) 2292--2295},
\href{http://arxiv.org/abs/astro-ph/9701231}{{\ttfamily arXiv:astro-ph/9701231
  [astro-ph]}}.

\bibitem{Zhang:2002xv}
B.~Zhang, Z.~G. Dai, and P.~M\'esz\'aros, ``{High-energy neutrinos from
  magnetars},'' \href{http://dx.doi.org/10.1086/377192}{{\em Astrophys. J.}
  {\bfseries 595} (2003) 346--351},
\href{http://arxiv.org/abs/astro-ph/0210382}{{\ttfamily arXiv:astro-ph/0210382
  [astro-ph]}}.

\bibitem{Metzger:2018uni}
B.~D. Metzger, T.~A. Thompson, and E.~Quataert, ``{A magnetar origin for the
  kilonova ejecta in GW170817},''
  \href{http://dx.doi.org/10.3847/1538-4357/aab095}{{\em Astrophys. J.}
  {\bfseries 856} no.~2, (2018) 101},
\href{http://arxiv.org/abs/1801.04286}{{\ttfamily arXiv:1801.04286
  [astro-ph.HE]}}.

\bibitem{Agasian:2001ym}
N.~O. Agasian and I.~Shushpanov, ``{Gell-Mann-Oakes-Renner relation in a
  magnetic field at finite temperature},''
  \href{http://dx.doi.org/10.1088/1126-6708/2001/10/006}{{\em JHEP} {\bfseries
  0110} (2001) 006},
\href{http://arxiv.org/abs/hep-ph/0107128}{{\ttfamily arXiv:hep-ph/0107128
  [hep-ph]}}.

\bibitem{Andersen:2012zc}
J.~O. Andersen, ``{Chiral perturbation theory in a magnetic background -
  finite-temperature effects},''
  \href{http://dx.doi.org/10.1007/JHEP10(2012)005}{{\em JHEP} {\bfseries 1210}
  (2012) 005},
\href{http://arxiv.org/abs/1205.6978}{{\ttfamily arXiv:1205.6978 [hep-ph]}}.

\bibitem{Hidaka:2012mz}
Y.~Hidaka and A.~Yamamoto, ``{Charged vector mesons in a strong magnetic
  field},'' \href{http://dx.doi.org/10.1103/PhysRevD.87.094502}{{\em Phys.
  Rev.} {\bfseries D87} no.~9, (2013) 094502},
\href{http://arxiv.org/abs/1209.0007}{{\ttfamily arXiv:1209.0007 [hep-ph]}}.

\bibitem{Luschevskaya:2014lga}
E.~V. Luschevskaya, O.~E. Solovjeva, O.~A. Kochetkov, and O.~V. Teryaev,
  ``{Magnetic polarizabilities of light mesons in $SU(3)$ lattice gauge
  theory},'' \href{http://dx.doi.org/10.1016/j.nuclphysb.2015.07.023}{{\em
  Nucl. Phys.} {\bfseries B898} (2015) 627--643},
\href{http://arxiv.org/abs/1411.4284}{{\ttfamily arXiv:1411.4284 [hep-lat]}}.

\bibitem{Bali:2017ian}
G.~S. Bali, B.~B. Brandt, G.~Endr\H{o}di, and B.~Gl{\"a}{\ss}le, ``{Meson
  masses in electromagnetic fields with Wilson fermions},''
\href{http://arxiv.org/abs/1707.05600}{{\ttfamily arXiv:1707.05600 [hep-lat]}}.

\bibitem{Fayazbakhsh:2012vr}
S.~Fayazbakhsh, S.~Sadeghian, and N.~Sadooghi, ``{Properties of neutral mesons
  in a hot and magnetized quark matter},''
  \href{http://dx.doi.org/10.1103/PhysRevD.86.085042}{{\em Phys. Rev.}
  {\bfseries D86} (2012) 085042},
\href{http://arxiv.org/abs/1206.6051}{{\ttfamily arXiv:1206.6051 [hep-ph]}}.

\bibitem{Avancini:2016fgq}
S.~S. Avancini, R.~L.~S. Farias, M.~Benghi~Pinto, W.~R. Tavares, and V.~S.
  Tim\'teo, ``{$\pi_0$ pole mass calculation in a strong magnetic field and
  lattice constraints},''
  \href{http://dx.doi.org/10.1016/j.physletb.2017.02.002}{{\em Phys. Lett.}
  {\bfseries B767} (2017) 247--252},
\href{http://arxiv.org/abs/1606.05754}{{\ttfamily arXiv:1606.05754 [hep-ph]}}.

\bibitem{Zhang:2016qrl}
R.~Zhang, W.-J. Fu, and Y.-X. Liu, ``{Properties of Mesons in a Strong Magnetic
  Field},'' \href{http://dx.doi.org/10.1140/epjc/s10052-016-4123-8}{{\em Eur.
  Phys. J.} {\bfseries C76} no.~6, (2016) 307},
\href{http://arxiv.org/abs/1604.08888}{{\ttfamily arXiv:1604.08888 [hep-ph]}}.

\bibitem{Mao:2017wmq}
S.~Mao and Y.~Wang, ``{Effect of Quark Dimension Reduction on Goldstone Mode in
  Magnetic Field},''
\href{http://arxiv.org/abs/1702.04868}{{\ttfamily arXiv:1702.04868 [hep-ph]}}.

\bibitem{GomezDumm:2017jij}
D.~G\'omez~Dumm, M.~Izzo~Villafa{\~{n}}e, and N.~N. Scoccola, ``{Neutral meson
  properties under an external magnetic field in nonlocal chiral quark
  models},'' \href{http://dx.doi.org/10.1103/PhysRevD.97.034025}{{\em Phys.
  Rev.} {\bfseries D97} no.~3, (2018) 034025},
\href{http://arxiv.org/abs/1710.08950}{{\ttfamily arXiv:1710.08950 [hep-ph]}}.

\bibitem{Coppola:2018vkw}
M.~Coppola, D.~G\'omez~Dumm, and N.~N. Scoccola, ``{Charged pion masses under
  strong magnetic fields in the NJL model},''
  \href{http://dx.doi.org/10.1016/j.physletb.2018.04.043}{{\em Phys. Lett.}
  {\bfseries B782} (2018) 155--161},
\href{http://arxiv.org/abs/1802.08041}{{\ttfamily arXiv:1802.08041 [hep-ph]}}.

\bibitem{Andersen:2012dz}
J.~O. Andersen, ``{Thermal pions in a magnetic background},''
  \href{http://dx.doi.org/10.1103/PhysRevD.86.025020}{{\em Phys. Rev.}
  {\bfseries D86} (2012) 025020},
\href{http://arxiv.org/abs/1202.2051}{{\ttfamily arXiv:1202.2051 [hep-ph]}}.

\bibitem{Fayazbakhsh:2013cha}
S.~Fayazbakhsh and N.~Sadooghi, ``{Weak decay constant of neutral pions in a
  hot and magnetized quark matter},''
  \href{http://dx.doi.org/10.1103/PhysRevD.88.065030}{{\em Phys. Rev.}
  {\bfseries D88} no.~6, (2013) 065030},
\href{http://arxiv.org/abs/1306.2098}{{\ttfamily arXiv:1306.2098 [hep-ph]}}.

\bibitem{Bali:2017yku}
G.~S. Bali, B.~B. Brandt, G.~Endr{\H{o}}di, and B.~Gl{\"a}{\ss}le, ``{Pion
  decay in magnetic fields},''
  \href{http://dx.doi.org/10.1051/epjconf/201817513005}{{\em EPJ Web Conf.}
  {\bfseries 175} (2018) 13005},
\href{http://arxiv.org/abs/1710.01502}{{\ttfamily arXiv:1710.01502 [hep-lat]}}.

\bibitem{Agashe:2014kda}
{\bfseries Particle Data Group} Collaboration, K.~A. Olive {\em et~al.},
  ``{Review of Particle Physics},''
\href{http://dx.doi.org/10.1088/1674-1137/38/9/090001}{{\em Chin. Phys.}
  {\bfseries C38} (2014) 090001}.

\bibitem{okun2013leptons}
L.~Okun, {\em Leptons and Quarks}.
\newblock North-Holland Personal Library. Elsevier Science, 2013.

\bibitem{schwartz2014quantum}
M.~Schwartz, {\em Quantum Field Theory and the Standard Model}.
\newblock Cambridge University Press, 2014.

\bibitem{Bhattacharya:2007vz}
K.~Bhattacharya, ``{Solution of the Dirac equation in presence of an uniform
  magnetic field},''
\href{http://arxiv.org/abs/0705.4275}{{\ttfamily arXiv:0705.4275 [hep-th]}}.

\bibitem{Bruckmann:2017pft}
F.~Bruckmann, G.~Endr\H{o}di, M.~Giordano, S.~D. Katz, T.~G. Kov\'acs,
  F.~Pittler, and J.~Wellnhofer, ``{Landau levels in QCD},''
  \href{http://dx.doi.org/10.1103/PhysRevD.96.074506}{{\em Phys. Rev.}
  {\bfseries D96} no.~7, (2017) 074506},
\href{http://arxiv.org/abs/1705.10210}{{\ttfamily arXiv:1705.10210 [hep-lat]}}.

\bibitem{Bali:2012zg}
G.~S. Bali, F.~Bruckmann, G.~Endr\H{o}di, Z.~Fodor, S.~D. Katz, and
  A.~Sch{\"a}fer, ``{QCD quark condensate in external magnetic fields},''
  \href{http://dx.doi.org/10.1103/PhysRevD.86.071502}{{\em Phys. Rev.}
  {\bfseries D86} (2012) 071502},
\href{http://arxiv.org/abs/1206.4205}{{\ttfamily arXiv:1206.4205 [hep-lat]}}.

\bibitem{Borsanyi:2010cj}
S.~Bors\'anyi, G.~Endr\H{o}di, Z.~Fodor, A.~Jakov\'ac, S.~D. Katz, {\em
  et~al.}, ``{The QCD equation of state with dynamical quarks},''
  \href{http://dx.doi.org/10.1007/JHEP11(2010)077}{{\em JHEP} {\bfseries 1011}
  (2010) 077},
\href{http://arxiv.org/abs/1007.2580}{{\ttfamily arXiv:1007.2580 [hep-lat]}}.

\bibitem{Kilcup:1986dg}
G.~W. Kilcup and S.~R. Sharpe, ``{A Tool Kit for Staggered Fermions},''
\href{http://dx.doi.org/10.1016/0550-3213(87)90285-9}{{\em Nucl. Phys.}
  {\bfseries B283} (1987) 493--550}.

\bibitem{Gimenez:1998ue}
V.~Gimenez, L.~Giusti, F.~Rapuano, and M.~Talevi, ``{Nonperturbative
  renormalization of quark bilinears},''
  \href{http://dx.doi.org/10.1016/S0550-3213(98)00582-3}{{\em Nucl. Phys.}
  {\bfseries B531} (1998) 429--445},
\href{http://arxiv.org/abs/hep-lat/9806006}{{\ttfamily arXiv:hep-lat/9806006
  [hep-lat]}}.

\bibitem{Gockeler:1998ye}
M.~G{\"o}ckeler, R.~Horsley, H.~Oelrich, H.~Perlt, D.~Petters, P.~E.~L. Rakow,
  A.~Sch{\"a}fer, G.~Schierholz, and A.~Schiller, ``{Nonperturbative
  renormalization of composite operators in lattice QCD},''
  \href{http://dx.doi.org/10.1016/S0550-3213(99)00036-X}{{\em Nucl. Phys.}
  {\bfseries B544} (1999) 699--733},
\href{http://arxiv.org/abs/hep-lat/9807044}{{\ttfamily arXiv:hep-lat/9807044
  [hep-lat]}}.

\bibitem{Skouroupathis:2008mf}
A.~Skouroupathis and H.~Panagopoulos, ``{Two-loop renormalization of vector,
  axial-vector and tensor fermion bilinears on the lattice},''
  \href{http://dx.doi.org/10.1103/PhysRevD.79.094508}{{\em Phys. Rev.}
  {\bfseries D79} (2009) 094508},
\href{http://arxiv.org/abs/0811.4264}{{\ttfamily arXiv:0811.4264 [hep-lat]}}.

\bibitem{Bali:2013kia}
G.~S. Bali, F.~Bursa, L.~Castagnini, S.~Collins, L.~Del~Debbio, B.~Lucini, and
  M.~Panero, ``{Mesons in large-N QCD},''
  \href{http://dx.doi.org/10.1007/JHEP06(2013)071}{{\em JHEP} {\bfseries 06}
  (2013) 071},
\href{http://arxiv.org/abs/1304.4437}{{\ttfamily arXiv:1304.4437 [hep-lat]}}.

\bibitem{Lubicz:2016xro}
V.~Lubicz, G.~Martinelli, C.~T. Sachrajda, F.~Sanfilippo, S.~Simula, and
  N.~Tantalo, ``{Finite-Volume QED Corrections to Decay Amplitudes in Lattice
  QCD},'' \href{http://dx.doi.org/10.1103/PhysRevD.95.034504}{{\em Phys. Rev.}
  {\bfseries D95} no.~3, (2017) 034504},
\href{http://arxiv.org/abs/1611.08497}{{\ttfamily arXiv:1611.08497 [hep-lat]}}.

\bibitem{Patella:2017fgk}
A.~Patella, ``{QED Corrections to Hadronic Observables},'' {\em PoS} {\bfseries
  LATTICE2016} (2017) 020,
\href{http://arxiv.org/abs/1702.03857}{{\ttfamily arXiv:1702.03857 [hep-lat]}}.

\bibitem{1991ApJ...383..745L}
D.~{Lai} and S.~L. {Shapiro}, ``{Cold equation of state in a strong magnetic
  field - Effects of inverse beta-decay},''
  \href{http://dx.doi.org/10.1086/170831}{{\em \apj} {\bfseries 383} (Dec.,
  1991) 745--751}.

\bibitem{Ferrer:2010wz}
E.~J. Ferrer, V.~de~la Incera, J.~P. Keith, I.~Portillo, and P.~L. Springsteen,
  ``{Equation of State of a Dense and Magnetized Fermion System},''
  \href{http://dx.doi.org/10.1103/PhysRevC.82.065802}{{\em Phys. Rev.}
  {\bfseries C82} (2010) 065802},
\href{http://arxiv.org/abs/1009.3521}{{\ttfamily arXiv:1009.3521 [hep-ph]}}.

\bibitem{Matese:1969zz}
J.~J. Matese and R.~F. O'Connell, ``{Neutron Beta Decay in a Uniform Constant
  Magnetic Field},''
\href{http://dx.doi.org/10.1103/PhysRev.180.1289}{{\em Phys. Rev.} {\bfseries
  180} (1969) 1289--1292}.

\bibitem{FassioCanuto:1970wk}
L.~Fassio-Canuto, ``{Neutron beta decay in a strong magnetic field},''
\href{http://dx.doi.org/10.1103/PhysRev.187.2141}{{\em Phys. Rev.} {\bfseries
  187} (1969) 2141--2146}.

\bibitem{Shinkevich:2004ja}
S.~Shinkevich and A.~Studenikin, ``{Relativistic theory of inverse beta decay
  of polarized neutron in strong magnetic field},''
  \href{http://dx.doi.org/10.1007/BF02898612}{{\em Pramana} {\bfseries 65}
  (2005) 215--244},
\href{http://arxiv.org/abs/hep-ph/0402154}{{\ttfamily arXiv:hep-ph/0402154
  [hep-ph]}}.

\bibitem{Giunti:2014ixa}
C.~Giunti and A.~Studenikin, ``{Neutrino electromagnetic interactions: a window
  to new physics},'' \href{http://dx.doi.org/10.1103/RevModPhys.87.531}{{\em
  Rev. Mod. Phys.} {\bfseries 87} (2015) 531},
\href{http://arxiv.org/abs/1403.6344}{{\ttfamily arXiv:1403.6344 [hep-ph]}}.

\end{thebibliography}\endgroup

\end{document}